\documentclass[draft]{agujournal2019}
\usepackage{url} %this package should fix any errors with URLs in refs.
\usepackage{lineno}
\usepackage{amsmath, color, algorithm2e, mathtools}
\usepackage[inline]{trackchanges} %for better track changes. finalnew option will compile document with changes incorporated.
\usepackage{soul}
\newcommand{\V}[1]{\mathbf{#1}} 
\newcommand{\B}{\mathbf{B}}

\newcommand{\gkeyll}{{\tt Gkeyll}}
\newcommand{\Alfven}{Alfv\'{e}n}
\DeclarePairedDelimiter{\norm}{\lVert}{\rVert}

\draftfalse

\journalname{Earth and Space Science (Technical Reports: Methods)}

\begin{document}

\title{Multi-Spacecraft Magnetic Field Reconstructions: A Cross-Scale Comparison of Methods}

\authors{T. Broeren\affil{1}, K. G. Klein\affil{2}, and J. M. TenBarge\affil{3}}
% ORCID for:
% Broeren [0000-0002-2649-020X]
% Klein [0000-0001-6038-1923]
% TenBarge [0000-0003-0143-951X]

\affiliation{1}{Department of Applied Mathematics, University of Arizona, Tucson, AZ, USA}
\affiliation{2}{Lunar and Planetary Laboratory, University of Arizona, Tucson, AZ, USA}
\affiliation{3}{Department of Astrophysical Sciences, Princeton University, Princeton, NJ, USA}

\correspondingauthor{Theodore Broeren}{Broeren@arizona.edu}

%% Keypoints, final entry on title page.

%  List up to three key points (at least one is required)
%  Key Points summarize the main points and conclusions of the article
%  Each must be 140 characters or fewer with no special characters or punctuation and must be complete sentences

\begin{keypoints}
\item We propose two new reconstruction methods that only require measurements of the magnetic field and bulk plasma velocity
\item The two new methods result in reconstructions with less error than previously employed Curlometer-based methods
\item The new methods preserve the statistical distribution of scale-dependent fluctuations from the ground-truth turbulence simulation
\end{keypoints}

\begin{abstract}
Space plasma studies frequently use in situ magnetic field measurements taken from many spacecraft simultaneously. A useful data product of these measurements is the reconstructed magnetic field in a volume near the spacecraft observatory. We compare a standard method of computing the magnetic field at arbitrary spatial points, the Curlometer, to two novel approaches: a Radial Basis Function interpolation and a time-dependent 2D inverse distance weighted interpolation scheme called Timesync. These three methods, which only require in situ measurements of the magnetic fields and bulk plasma velocities at a sparse set of spatial points, are implemented on synthetic data drawn from a time-evolving numerical simulation of plasma turbulence. We compare both the topology of the reconstructed field to the ground truth of the simulation and the statistics of the fluctuations found in the reconstructed field to those from the simulated turbulence. We conclude that the Radial Basis Function and Timesync methods outperform the Curlometer in both the topological and statistical comparisons.
\end{abstract}

\section*{Plain Language Summary}
Spacecraft are often used to measure things, such as magnetic fields, which can only be known at a single point, the point where a spacecraft is located. When multiple spacecraft are deployed to measure the magnetic field as a group, we get an understanding of how the magnetic field changes by comparing the measurements made by each of the spacecrafts against one another. When multiple spacecraft are taking measurements, scientists also use mathematical techniques to estimate the magnetic field at points in space where a spacecraft is not located. The process of using a small number of magnetic field measurements, taken from spacecraft, to estimate the value of the magnetic field everywhere is called magnetic field reconstruction. This work has compared the effectiveness of three of these magnetic field reconstruction techniques: one established and two novel. We conclude that the two novel methods are more accurate than the established mathematical tool commonly used to reconstruct magnetic fields.

\section{Introduction}
Plasmas are ubiquitous throughout the universe and difficult to maintain in a laboratory. Therefore, space is a natural laboratory for the study of plasmas, in particular the solar wind. Since magnetic fields and moving charged particles are self-consistently coupled, a critical component of plasma studies is knowledge of the magnetic field structure. Often, the most reliable magnetic field observations can only be made with in situ measurements. The desire to measure the spatial structure of these fields has led to multi-spacecraft missions such as Cluster \cite{Escoubet:1997}, MMS \cite{Burch:2016}, and HelioSwarm \cite{Klein:2023}, which take measurements over distributed observatories. These simultaneous multi-point experiments spurred the development of multi-point analysis techniques to estimate the value of the magnetic field in a region near the constellation of spacecraft \cite{Paschmann:1998, Paschmann:2008}. 

The multi-point analysis techniques under consideration in this work can be broadly classified as sparse data interpolation methods. We compare a standard method of estimating the magnetic field at an unknown point, the Curlometer method (\S \ref{ssec:curl}) \cite{Dunlop:1988}, to two alternatives. For each method, we use timeseries data drawn from synthetic multi-spacecraft measurements of a numerical simulation of turbulence to reconstruct the 3D magnetic field (see setup in \S \ref{ssec:timeseries}).

The fidelity of multi-point spacecraft analysis techniques are typically studied using in situ spacecraft data \cite{Narita:2013, Denton:2020} or a snapshot in time of a turbulent plasma simulation \cite{Denton:2022}. As we wish to capture the effects of time-varying plasmas, we use synthetic spacecraft data drawn from a time dependent simulation of plasma turbulence as inputs for our study of three different magnetic field reconstruction methods; the details of the simulation are in \S \ref{ssec:turb_sim}.

The first alternative method, Radial Basis Function (RBF) interpolation (\S \ref{ssec:rbf}), can be interpreted as a simple single-layer neural network without regularization. This method of interpolation and approximation is widely used in many other areas of science such as time series prediction, 3D computer graphics, and control of chaotic nonlinear systems.

The second alternative, which we refer to as Timesync (\S \ref{ssec:TS}), is an inverse distance weighted scheme that tackles the 3D problem as a series of 2D planes. These 2D planes are stitched together (as a function of time) to build information in the third spatial dimension parallel to the relative motion between the spacecraft and plasma. We treat the most densely sampled spatial direction independently to minimize intra-spacecraft contamination at small scales.

We use a variety of metrics (described in \S \ref{ssec:evaluation_methods}) to quantify the goodness of the three reconstruction methods. This quantification is a necessity because turbulent magnetic fields have features that we wish to preserve at a variety of spatial scales simultaneously. Additionally, each metric we apply has its own intrinsic limitations and faults, requiring that we consider multiple techniques to verify that our analysis is accurate and complete. We perform the quantification of the three analysis methods using these metrics on three different multi-spacecraft configurations (\S \ref{ssec:sc_config}) to characterize the influence of both the number of spacecraft and shape of configuration. The results of these analyses can be found in \S \ref{sec:results}, followed by our conclusions in \S \ref{sec:conclude}.

\section{Reconstruction Methods}
We select three distinct methods of reconstructing magnetic fields from arbitrary spacecraft configurations. As both measurements of magnetic fields and bulk plasma velocities are commonly made in situ by space plasma missions, we chose to compare methods that only require in situ measurements of magnetic fields and plasma velocities. We test our methods using a 3D time-varying numerical simulation of plasma turbulence. Synthetic spacecraft measurements are constructed by moving a fixed spacecraft configuration through the simulated plasma in the $\hat{x}$ direction at a constant $320$ km/s. It is important to note that the plasma is evolving in time as the spacecraft are moving through it in the $\hat{x}$ direction, and the value of spacecraft velocity is selected to approximate the solar wind velocity \cite[Table 1]{Chasapis:2020}.

Our two new reconstruction methods assume that the spatial evolution of the plasma dominates its temporal evolution \cite{taylor:1938} over short intervals in time. This assumption allows us to expand our sparse set of measurement points in space (over $N$ spacecraft) and time (over timeseries of length $T$), enhancing the reconstruction methods' accuracy.

\subsection{Curlometer}
\label{ssec:curl}
The Curlometer technique was originally developed for the Cluster mission \cite{Dunlop:2002}, but has been since been adapted for the multi-spacecraft missions MMS \cite{Burch:2016}, THEMIS \cite{Angelopoulos:2008}, and Swarm \cite{Dunlop:2021}. While this method was originally formulated to estimate current density within a configuration of four spacecraft, it can also be used to reconstruct the magnetic field near a spacecraft configuration. This reconstruction method is the theoretical basis for the FOTE model \cite{Fu:2015} and the LB-3D model \cite{Denton:2020} (which adds the additional constraint that $\nabla \cdot B = 0$). The method can be written as a first-order Taylor series approximation
\begin{equation}
B_m(\mathbf{r}^{(i)}) = B_m(\mathbf{r}) + \sum_{k \in \{x,y,z\}} \partial_k B_{m} \left(r^{(i)}_k - r_k\right)  \hspace{0.7cm} \forall  \hspace{0.2cm} i \in \{1,2,3,4\},m \in \{x,y,z\}.  \label{eqn:curl_O1} 
\end{equation}
This method gives us a system of twelve equations and twelve unknowns where $B_m(\mathbf{r}^{(i)})$ is the measured $m^{th}$ component of $\B$ at the $i^{th}$ spacecraft position $\mathbf{r}^{(i)}$, $B_m(\mathbf{r})$ is the computed $m^{th}$ component of $\B$ at the reconstructed point $\mathbf{r}$, and $\partial_k B_{m}$ is the computed $k$ spatial derivative of the $m^{th}$ component of $\B$ at $\mathbf{r}$.

Our previous work has shown that the Curlometer can be extended to configurations of more than four spacecraft by summing over the combinatorially large number of four spacecraft subsets that exist within more numerous configurations \cite{Broeren:2021}. Also in that work, we showed that the performance of a four-spacecraft subset (a tetrahedron) was related to its characteristic size, $L$, elongation, $E$, and planarity, $P$. Elongation and planarity ($E,P \in [0,1]$) are measures of dissimilarity between the semi-axis lengths of a representative ellipsoid, derived from the positions of the spacecrafts \cite[Chapter 16.3]{Paschmann:1998}. From $E$ and $P$, we define a single composite shape 'goodness' parameter
\begin{equation}
    \chi = \sqrt{E^2 + P^2} \in [0,\sqrt{2}] .\label{eqn:chi}
\end{equation}

\begin{figure}
\centering
\textbf{Reconstruction Methods}\par\medskip
\includegraphics[width=1\textwidth]{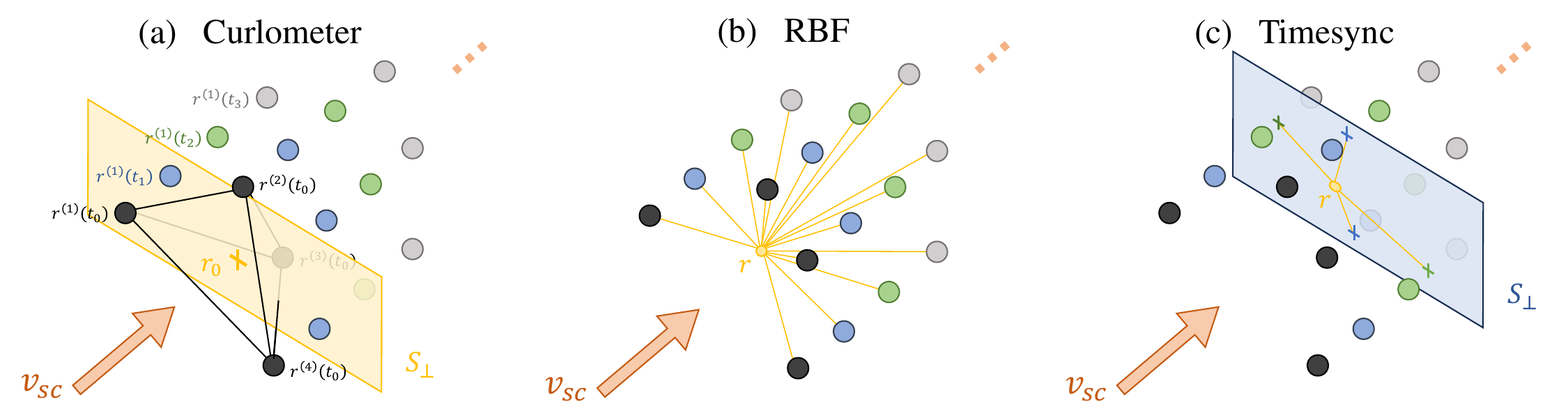}
\caption{An illustration of how spacecraft measurements are combined for the three reconstruction methods. The Curlometer considers $N$ measurements for each reconstructed point, one from each spacecraft at time $t_i$ to reconstruct the plane of points $S_{\perp}$ that intersects the configuration's barycenter $r_0$. The RBF method weighs all measurements made by the $N$ spacecraft at all $T$ points in time. The Timesync method also considers $N$ measurements at a time, one from each spacecraft. However, for this method, the times at which these samples are taken is not necessarily the same for all of the $N$ spacecraft.}
\label{fig:All_methods}
\end{figure}

We use $\chi$ to define thresholds in selecting tetrahedra that are well shaped, with lower values being better due to a more spherical distribution of spacecraft. For a configuration of $N$ spacecraft, we compute the $\chi$ value for each of the 
\begin{equation}
    \begin{pmatrix} N \\ 4 \end{pmatrix} = \frac{N!}{4!(N-4)!}
\end{equation}
tetrahedra. We use the tetrahedra that have $\chi < 0.6$ to estimate the magnetic field for all points on the $yz$-plane passing through the barycenter, 
\begin{equation}
    \mathbf{r}_0 = \frac{1}{N} \sum_{i=1}^N \mathbf{r}^{(i)},
\end{equation}
of the overall configuration. We then compute the component-wise median of these different values to give us our final estimation of the magnetic field on the $yz$-barycentric plane. This process is then repeated independently for every time step of the analysis (see Figure \ref{fig:All_methods} for setup). We note that we use the median instead of the mean of the estimates because it can be shown analytically that Curlometer estimates, as well as linear combinations of them, do not preserve scale-dependent fluctuations, while the median of a group of estimates does not have such a theoretical limitation (see \S \ref{appendix:sec:curl_scale}).

This method does not assume Taylor's Hypothesis is valid. We tested a version of this method where Taylor's Hypothesis was assumed to be true, which expanded the number of possible tetrahedra to $\begin{pmatrix} NT \\ 4 \end{pmatrix}$. However, this implementation did not improve results but drastically increased computational complexity.

For this method, we solve Eqn \ref{eqn:curl_O1} exactly, and therefore do not enforce that the reconstructed magnetic field be solenoidal ($\nabla \cdot \mathbf{B} = 0$). Previous works have applied this constraint by adding an additional equation and solving the resulting $13\times 12$ system in a least squares sense \cite{Denton:2020}. However, we found that this constraint did not appear to improve the accuracy of the reconstructed fields.

\subsection{RBF Interpolation}
\label{ssec:rbf}
Radial Basis Function (RBF) interpolation was first introduced as a method of reconstructing topological surfaces from scattered data points \cite{Hardy:1971}. It was subsequently applied to the fields of geophysics, surveying, photogrammetry, remote sensing, signal processing, geography, hydrology, and elliptic/parabolic/hyperbolic PDEs \cite{Hardy:1988}. Recently, it has been adopted by the computational community, as it can be interpreted and used as a simple neural network \cite{Broomhead:1988, Elanayar:1994, Schilling:2001}.

A radial basis function $\varphi(\norm{\mathbf{r}^{(i,j)} - \mathbf{r}})$ is a function whose value depends only on the distance between the input and some fixed point, called a center, $\mathbf{r}^{(i,j)}$. We use a weighted sum of these RBFs to interpolate $\B$ to arbitrary positions $\mathbf{r}$ using measurements made at the $N$ spacecraft locations over all $T$ timesteps, $\mathbf{r}^{(i,j)}$ (i.e., the spacecraft positions are used as centers)
\begin{equation}
    B_m(\mathbf{r}) = \sum_{i=1}^{N} \sum_{j=1}^{T} a_{i,j} \varphi(\norm{\mathbf{r}^{(i,j)} - \mathbf{r}}) \hspace{0.5cm} \forall m \in \{x,y,z\}.\label{eqn:RBF}
\end{equation}
To perform this interpolation, we must pick a functional form for $\varphi$ and we must learn the values of the constant weights $a_{i,j}$. The estimated values of these constants are learned through weighted linear regression, using the measured values of magnetic field and spacecraft positions as a dataset to be fit to. This method is easily implemented in Python using the Scipy interpolate packages \textit{Rbf} and \textit{RBFInterpolator} \cite{SciPy:2020, Fasshauer:2007}.

This method can also be viewed as learning the weights of an RBF neural network without regularization, so that the data fits the learned equation/network exactly. Importantly, our results are not dependent on a user selected training dataset. Rather, the sparse dataset that we wish to interpolate is used for both the computation of the constant weights and the interpolation between them.

Several smooth RBF kernel functions are used in the literature including Gaussian, multiquadric, and inverse quadric \cite{Fasshauer:2007}. We tested these three options and found that the multiquadric kernels performed the best. The multiquadric RBF kernel has the form
\begin{equation}
    \varphi(\norm{\mathbf{r}^{(i,j)}-\mathbf{r}}) = \sqrt{1 + \left(\norm{\mathbf{r}^{(i,j)}-\mathbf{r}}/\sigma \right)^2}. \label{eqn:MQ_RBF}
\end{equation}
Here, $\sigma$ is a tunable hyper-parameter that is proportional to the radius of influence that measured data points have on the interpolation.

To select an optimal value of $\sigma$, previous studies have proposed a “leave-one-out” cross-validation algorithm \cite{Rippa:1999}. This algorithm identifies a cost function which numerically imitates the behavior of the RMS error of the interpolation itself. By finding the $\sigma$ that minimizes this cost function, they were also able to approximate the $\sigma$ that would minimize the RMS error of RBF interpolation. We note that while this method has been analyzed and modified by others \cite{Scheuerer:2010, Ghalichi:2022}, we use this method in its original form to select an appropriate $\sigma$. This method allows us to have a $\sigma$ parameter that is dependent on spacecraft spatial configuration, but independent of the data that is being collected.

To increase the base of known points for this method, we use all data collected from the $N$ spacecraft at all $T$ points in time (see Figure \ref{fig:All_methods}), providing us with $NT$ RBF centers (and learned coefficients) in the sum in Eqn \ref{eqn:RBF}. This approach may seem like we are invoking Taylor's Hypothesis over the entire interval of the time series. However, as the distance from a spacecraft measurement to a reconstructed point becomes much larger than $\sigma$, the weight assigned to that measurement decreases to zero in the RBF sum. This weighting means measurements made farther apart in space and/or time will become less correlated than those made more closely in distance/time. Therefore, in the RBF method, Taylor's hypothesis is being effectively invoked over a distance of $\sim \sigma$.

A select few previous works have utilized a modified RBF method to reconstruct the magnetosphere or magnetic structures measured by spacecraft \cite{andreeva:2016, chen:2019}. These approaches differed from ours, as they demonstrated that the RBF method can be adapted to enforce a divergence-free field by decomposing $\B$ into toroidal and poloidal components, each of which is learned separately via RBF interpolation. We do not use this modified approach because the decomposition method intrinsically computes gradients of the magnetic field, which we wish to avoid in order to maintain accuracy.

\subsection{Timesync}
\label{ssec:TS}
It is often the case that magnetic field measurements are taken with high temporal cadence, but with sparse spatial sampling. In the solar wind, a typical bulk plasma is flowing at approximately 320 km/s \cite{Chasapis:2020}. For multi-spacecraft missions, this means observatory measurements will be uniformly dense along the direction of advection (defined by the sampling rate) and non-uniformly sparse along the 2 perpendicular directions (defined by the spacecraft separations). This asymmetry in sampling density has been shown to be non-optimal for RBF interpolation schemes \cite{DeMarchi:2005}. To address this issue, we define our own method, called Timesync, which is better suited to handle asymmetrically sampled data.

First, we use our knowledge of the spacecraft velocity and measurements of the local average plasma velocity to identify the spacecraft's direction of travel in the stationary plasma frame. For the rest of this work, we will refer to this as velocity $\mathbf{v}_{sc}$. To reconstruct an arbitrary point, we treat the data in the $\mathbf{v}_{sc}$ direction separately from the 2D perpendicular plane $S_{\perp}$. 

In the $\mathbf{v}_{sc}$ direction, we use a 1D interpolation to produce each spacecraft's timeseries at an arbitrary time cadence. For this work, we have used nearest neighbor interpolation, but in principle many 1D methods (polynomial spline, windowed averaging, etc.) could be used. At each point in time that we wish to reconstruct the magnetic field, we identify the plane that is perpendicular to $\mathbf{v}_{sc}$ and also intersects the reconstructed point, $S_{\perp}$. 

We perform a 2D inverse distance weighted (IDW) interpolation on each $S_{\perp}$ plane independently to estimate values spanning this surface. There is only one measurement per spacecraft that is taken into account for these 2D reconstructions: each 1D-interpolated measurement from the spacecraft's timeseries gives one value on $S_{\perp}$ (see Figure \ref{fig:All_methods}).

Inverse distance weighted interpolation is another very common interpolation method for irregularly spaced multi-dimensional data \cite{Shepard:1968, Lu:2008}. The method is defined as
\begin{equation}
\B(\mathbf{r}) = \begin{cases}
\frac{\sum_{i=1}^N w_i(\mathbf{r}) \B(\mathbf{r}^{(i)})}{\sum_{i=1}^N w_i(\mathbf{r})} &\text{if $\norm{\mathbf{r}^{(i)}-\mathbf{r}}\neq 0$}\\
\B(\mathbf{r}^{(i)}) &\text{if $\norm{\mathbf{r}^{(i)}-\mathbf{r}} = 0$}
\end{cases}
\end{equation}
where 
\begin{equation}
    w_i = \frac{1}{\norm{\mathbf{r}^{(i)}-\mathbf{r}}^2}.
\end{equation}
Using this combined nearest neighbor/IDW approach, we are not reliant on magnetic field data smoothing, such as we would be if using the LB-3D method \cite{Denton:2022} or a Grad-Shafranov based approach \cite{Sonnerup:2006}. This approach also has the benefit that it contains no hyper-parameters which need to be tuned, such as the $\sigma$ term in the RBF interpolation scheme. This method assumes that the plasma does not evolve in the time it takes a leading and trailing spacecraft to pass through the same transverse plane of plasma. Therefore, this method assumes Taylor's Hypothesis holds over the spacecraft configurations characteristic size $\sim L$.

\section{Comparison Methodology}
To evaluate each of these methods in a scenario that mirrors that found from in situ spacecraft measurements of solar wind plasmas, we use synthetic data generated from a time-varying numerical simulation of plasma turbulence. We use the simulation to extract synthetic magnetic field timeseries for each of the spacecraft from three selected configurations. We then implement each of the three reconstruction methods on the data and compare these reconstructions to the ground truth of the underlying plasma simulation.

\subsection{Turbulence Simulation}
\label{ssec:turb_sim}
We utilize the magnetic fields from a fully developed turbulence simulation performed within the \gkeyll\ simulation framework \cite{Hakim:2006,Wang:2015,Wang:2020}. This turbulence simulation is designed to represent plasma behavior in the pristine solar wind at 1AU.

We use the five moment ($n_s, \V{u}_s, p_s$), two fluid ($s=p,e$) plasma model to evolve a proton-electron plasma with a reduced mass ratio of $m_p / m_e = 100$, a temperature ratio of $T_p / T_e = 1$, \Alfven\ velocity of $v_{A} /c =  B_0/\sqrt{\mu_0 n_p m_p c^2} =0.02$, plasma beta of $\beta_p = 2\mu_0 n_p T_p/B_0^2 = 1$, and adiabatic index $\gamma = 5/3$. We employ an elongated domain $L_x = L_y = 0.2 L_z = 100 \pi \rho_p$ with spatial resolution of $448^3$. Lengths are normalized to the proton gyroradius $\rho_p = v_{tp}/\Omega_p$, the ratio of the proton thermal speed $v_{tp} = \sqrt{2T_p/m_p}$ and the proton cyclotron frequency $\Omega_p = q_p B/m_p$. We choose a uniform background density and magnetic field, $\V{B}_0 = B_0 \hat{\V{z}}$, and an initial turbulence amplitude $\delta B / B_0 = 0.2$ to satisfy critical balance.

To insert the selected spacecraft configurations with separations in physical units into the dimensionless simulation, we set $n_e=0.2829$ cm$^{-3}$, which defines our proton gyroradius $\rho_p = 100$ km. To initialize the simulation, it is run for one \Alfven\ crossing time, $t_A = 1500 / \Omega_{p}$, at which point the turbulence has fully developed and reached a steady state. We then save a series of 225 samples of the 3D plasma cube in time. A more detailed description of this plasma simulation can be found in \S 2.3.2 of our Curlometer extension paper \cite{Broeren:2021}, where we used the resulting magnetic fields from a single instant in time.

\subsection{Spacecraft Configurations}
\label{ssec:sc_config}
We define three configurations of spacecraft which we will use in this analysis, configurations A, B and C. All three of these configurations have the same characteristic size $L=2000$km. However, they vary in number of spacecraft and shape of configuration.

\textbf{Configuration A} is a four-spacecraft configuration. It has an elongation and planarity of $0.05$ and $0.01$ respectively. This configuration represents the near-tetrahedral formation that is often desired in missions such as MMS or Cluster.

\textbf{Configuration B} is a nine-spacecraft configuration. It has an overall elongation and planarity of $0.02$ and $0.01$ respectively. Its relative spacecraft positions, which were randomly generated for a previous study \cite[\S 2.4.1]{Broeren:2023}, are approximately uniformly distributed throughout a cube. This configuration allows us to vary the number of spacecraft, while keeping the approximate shape of the configuration fixed in comparison with configuration A.

\textbf{Configuration C} is also a nine-spacecraft configuration. It has an overall elongation and planarity of $0.34$ and $0.67$ respectively. This configuration was drawn from the NASA HelioSwarm mission (hour 150 of the C1 Phase-B Design Reference Mission) scaled to have a characteristic size of $2000$km \cite{levinson:2021}. This third configuration allows us to vary the shape of the configuration, while keeping the number of spacecraft constant in comparison with configuration B. It will also inform us of the typical HelioSwarm configuration performance.

\subsection{Synthetic Timeseries Generation}
\label{ssec:timeseries}
We extract 20 synthetic timeseries of data out of the turbulence simulation from spatially-disjoint regions of the global plasma simulation. Each timeseries is generated under the assumption that the spacecraft configuration is fixed and that the solar wind, with velocity 320 km/s, is flowing in the $-\hat{x}$ direction. Each timeseries taken from the time-varying plasma consists of $T=225$ samples taken at a 4 Hz sampling cadence, which corresponds to $\Omega_{p}=0.2\bar{6}$ sec.

\begin{figure}
\centering
\textbf{$B_x$ Component of Example Reconstruction}\par\medskip
\includegraphics[trim={0.1in 0.7in 0.1in 0.6in}, clip, width=.98\textwidth]{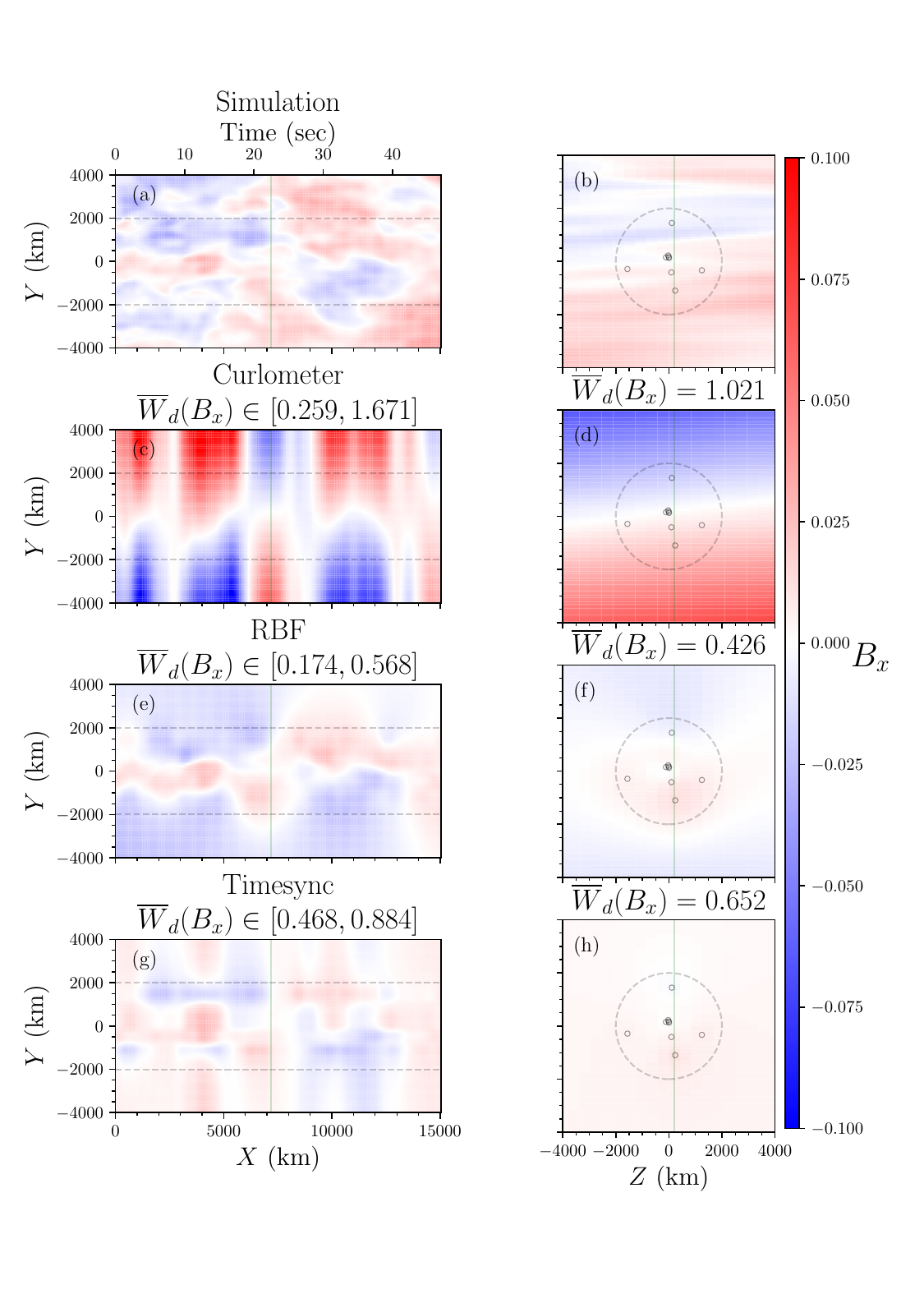}
\caption{An example reconstructed magnetic field around spacecraft configuration C (small black circles are spacecraft positions $\mathbf{r}^{(i)}$). We show the $x$ component of the simulated magnetic field (top row) vs the $x$ component of the reconstructed fields (subsequent rows) using all three reconstruction methods using contour plots. The green lines indicate which slice in the 3D volume is pictured in the adjacent panel. The relative Wasserstein distance, $\overline{W}_d$ (Eqn \ref{eqn:Wd_tilde}), for each of these example reconstructed $yz$-planes is computed and listed for reference on the right panels. We also list the ranges of $\overline{W}_d$ values observed over the entire interval above the left panels ($10\%$ and $90\%$ percentile values). The black dashed lines and circle correspond to the cylindrical region within one characteristic size $L$ of the spacecraft configurations barycenter $\mathbf{r}_0$ where $\overline{W}_d$ is computed.}
\label{fig:Bx_compare}
\end{figure}

\begin{figure}
\centering
\textbf{Streamlines of Example Reconstruction}\par\medskip
\includegraphics[trim={0.1in 0.7in 0.85in 0.6in}, clip, width=.98\textwidth]{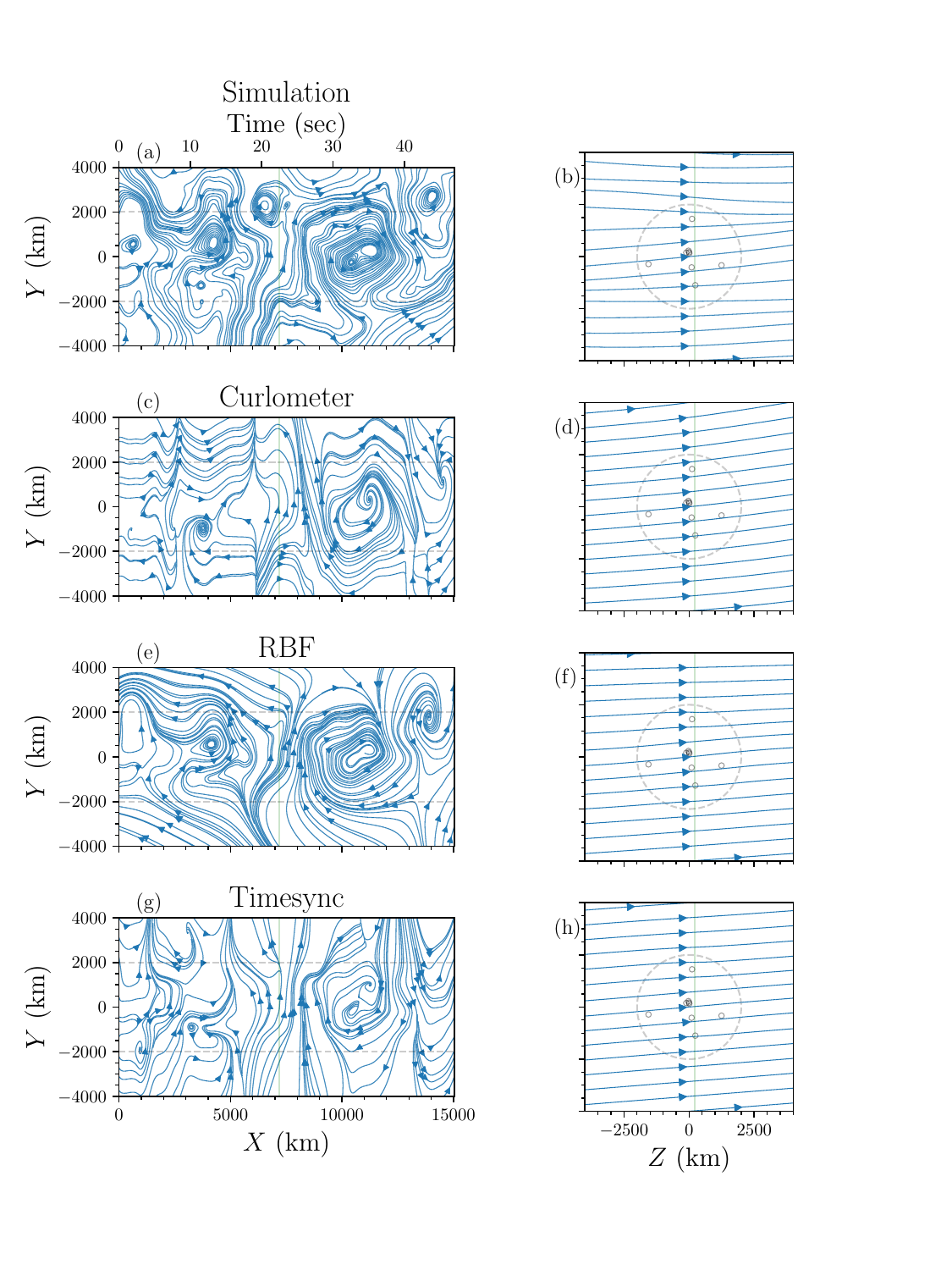}
\caption{Streamlines of the reconstructed magnetic field around spacecraft configuration C shown in Figure \ref{fig:Bx_compare}, both for the simulated magnetic field (top row) and the reconstructed fields using all three reconstruction methods (subsequent rows). }
\label{fig:streamplot_compare}
\end{figure}

In a $4L\times 4L$ square centered at the overall spacecraft configurations barycenter, we also extract a 3-dimensional grid of magnetic field vector values which will be compared to the reconstructed values at the same spatial positions. Trilinear interpolation is used to get values of $\B$ at non-grid points in the simulation volume. The spatial resolution of the reconstructed grid is $80 \text{ km} \times 81.63 \text{ km} \times 400$ km in space, which corresponds to a grid resolution of $225\times 99\times 20$ in the $x(t) \times y \times z$ directions. Using 20 independent realizations, this gives us an ensemble of nearly 9 million magnetic field vectors for each spacecraft configuration and reconstruction method combination. We visualize an example reconstructed field using each method in Figures \ref{fig:Bx_compare} and \ref{fig:streamplot_compare} and compare the reconstructions to the simulation.

\subsection{Evaluating Reconstruction}
\label{ssec:evaluation_methods}
We apply each reconstruction technique to the timeseries data and compare the reconstructed magnetic field to the ground-truth magnetic field, gathered from the time-varying turbulence simulation. This section describes the methods used to quantify the accuracy of the reconstructed magnetic fields.

\subsubsection{Point-Wise Error}
To compare the reconstructed magnetic field to the true field drawn from the simulation, we define the vectoral point-wise error as
\begin{equation}
    \text{Error} = 100 \frac{\norm{\mathbf{B}_{recon} - \mathbf{B}_{sim} }}{\norm{\mathbf{B}_{sim}}}. \label{eqn:Error_B}
\end{equation}
We compute this quantity at all spatial locations in the volume described in \S \ref{ssec:timeseries} for the 20 sets of synthetic plasma measurements. We then take the arithmetic mean of this error over those 20 iterations to find the error spatially with respect to a fixed spacecraft configuration. Finally, we use this average spatial error to quantify the fraction of the volume within a $2L$ radius of the spacecraft's barycenter that is reconstructed with a desired level of accuracy.

This point-wise error method is prone to suffer from the double penalty effect \cite{Farchi:2016}. If a magnetic field structure is present, but translated in position, then this effect penalizes the point-wise error both where the magnetic field structure should be and where the magnetic field structure is located in the reconstruction. To fully understand the topology of our results, we compliment the point-wise error analysis with the Wasserstein distance metric.

\subsubsection{Wasserstein Distance}
The Wasserstein distance is a measure of similarity between two probability distribution functions (pdfs) \cite{Givens:1984}. This measure differs from the comparison done in the Kolmogorov–Smirnov \cite{Berger:2014} or Anderson-Darling \cite{Scholz:1987} tests, as those measure the difference in the cumulative distribution functions, $F$ and $G$. The Wasserstein distance measures the differences in the underlying set of random variables themselves.

The Wasserstein distance computes the difference between two pdfs, $f$ and $g$. If we let $f$ have samples $f_i$ and $g$ have samples $g_i$, then the Wasserstein distance is defined as
\begin{equation}
    W_d(f,g) = \inf_{\pi} \left( \frac{1}{n} \sum_{i=1}^n \norm{f_i - g_{\pi(i)}}^p  \right)^{1/p} , \label{eqn:wass_dist}
\end{equation}
where the infimum is over all permutations $\pi$ of the $n$ samples of each pdf. In this work, we use the case of $p=1$, also known as the Earth movers distance. Heuristically, this distance is the minimum amount of 'work' to turn one distribution into the other, weighing the samples by the $L^p$-norm distance. 

The result of this computation is a single non-negative scalar that is zero if the distributions match exactly, and large if they do not. This metric has been used to recognize textures and patterns \cite{Rubner:2000}, to validate atmospheric dispersion models using real data \cite{Farchi:2016}, and to compare features present in 2D and 3D vector fields \cite{Lavin:1998, Batra:1999}. It is easily computed in Python using the Scipy stats package \textit{wasserstein distance} \cite{SciPy:2020, ramdas:2017}. 

We define our own error metric, the relative Wasserstein distance $\overline{W}_d$, to compare the topological similarity of two magnetic fields
\begin{equation}
    \overline{W}_d\left(B_{recon}, B_{sim}\right) = \frac{W_d\left(B_{recon}, B_{sim}\right)}{W_d\left(\widetilde{B}_{sim}, B_{sim}\right)}. \label{eqn:Wd_tilde}
\end{equation}
The relative Wasserstein distance is the Wasserstein distance between the reconstructed magnetic field $\B_{recon}$ and simulated (true) magnetic field $\B_{sim}$, normalized by the distance between the simulated magnetic field and the median value of the simulated magnetic field $\widetilde{B}_{sim}$. We compute this value separately for the $x$, $y$, and $z$ components of the magnetic field on every $yz$-plane in the reconstruction and simulation. 

The normalized quantity $\overline{W}_d$ represents the quality of the reconstructed field compared to a field which is approximated as a constant value over the same region. The constant value chosen, $\widetilde{B}_{sim}$, is the median value of the simulated magnetic field over the region of comparison. Therefore, a value of $\overline{W}_d = 0.25$ is interpreted as the reconstructed field being four times as good as any which can be approximated as a constant over the same region. It follows that any value of $\overline{W}_d$ greater than unity means the field was reconstructed with worse accuracy than what could be achieved using a constant value.

\subsubsection{Statistical Fluctuation Distribution}
To compare the turbulent fluctuations in the reconstructed magnetic field to those found in the turbulence simulation itself, we analyze the distributions of magnetic field fluctuations 
\begin{equation}
    |B_m(\mathbf{r}) - B_m(\mathbf{r + \delta})| \label{eqn:B_dist}
\end{equation}
for different lag distances $\delta$ in the $x$ and $y$ directions. Recall that the simulation is elongated and has a background magnetic field oriented along the $z$ direction. Therefore, we are analyzing two directions, $x$ and $y$, that are perpendicular to the mean field, one which is aligned with the direction of flow ($x$) and one which is not ($y$).

\subsubsection{Structure Function}
We also compute and plot the structure function
\begin{equation}
    S_n(\delta) = \langle |B_m(\mathbf{r}) - B_m(\mathbf{r + \delta})|^n \rangle , \label{eqn:S_1}
\end{equation}
where $n$ is the order of the structure function. We calculate $S_1$ from the reconstructed magnetic fields with lags $\mathbf{\delta}$ in the interval $[80,1000]$ km. We compare this to $S_1$ drawn from the turbulence simulation itself. We analyze lags $\delta_x$ and $\delta_y$, which are oriented in the $x$ or $y$ directions. 

We also find the scaling exponent, $\zeta(n)$, of the structure functions by fitting the equation
\begin{equation}
    S_n \propto \delta^{\zeta(n)}
\end{equation}
for values of $n \in \{1,2,3,4\}$. This scaling exponent ignores the deviations in total magnitude of $S_n$ that may be present in the reconstructed field data, focusing solely on the scaling with order $n$.

\section{Results}
\label{sec:results}
In Figures \ref{fig:Brecon_error_yz} through \ref{fig:zeta}, we show results for the three example multi-spacecraft configurations A, B, and C. These configurations, which correspond to both well-shaped and realistic geometries, range from $N=4$ spacecraft to $N=9$ spacecraft. 

For the Curlometer reconstruction method, we use a shape threshold of $\chi \leq 0.6$ for subsets within the $N=9$ spacecraft configurations. This threshold results in 34 of 126 and 9 of 126 tetrahedra passing the shape criteria for the B and C configurations respectively. For the RBF method, we use values of $\sigma = 213.0$, $303.9$, and $337.2$ km for configurations A, B, and C respectively. These $\sigma$s were computed using the data-driven cross-validation algorithm \cite{Rippa:1999}.

\subsection{Computational Times}
The Curlometer method is about twice as fast as the RBF and Timesync methods for spacecraft configuration A with $N=4$. However, it has a larger computational time than the RBF and Timesync for configurations B and C with $N=9$. This slowdown is because the number of tetrahedra scales super-exponentially with the number of spacecraft. The Curlometer is also the only method where computational time depends on the specific configuration shape, not just number of spacecraft. This dependence is because a constant tetrahedral shape goodness threshold in $\chi$ will select/disregard a different number of tetrahedra for different configurations of $N$-spacecraft. The computational time of the Curlometer method also scales linearly with the number of time samples, $T$.

The overall computational time of the RBF and Timesync methods is comparable: within 15\% for each configuration. The RBF and Timesync methods also both appear to scale approximately linearly with the number of spacecraft in the configuration, $N$, as well as the number of time samples in each of their respective timeseries, $T$. This scaling is obviously much more advantageous for future many-spacecraft missions, such as the proposed MagneToRE, which could contain huge numbers of spacecraft \cite{Maruca:2021}.

\subsection{Divergence of Reconstructions}
Because none of the three reconstruction methods strictly enforce that the reconstructed magnetic field be divergence free, we use this condition as one of our tests of quality. From the numerical simulation, we computed the baseline level of total divergence 
\begin{equation}
    \epsilon = \int_D |\nabla \cdot \B| dx^3 \label{eqn:eps_divB}
\end{equation}
of the field in a cylindrical region $D = \{(x,y,z):\sqrt{(y-y_0)^2 + (z-z_0)^2} \leq L\}$, where $\mathbf{r}_0 = (x_0(t), y_0, z_0)$ is the barycenter of each spacecraft configuration. To compute this using our finite number of grid points, we estimate the derivatives via a central difference scheme and sum the absolute values of divergence at all grid points which fall in the region $D$. 

We compute the median value of total divergence over the same region of space for the magnetic fields reconstructed using each of the three reconstructed methods and compare the value to $\epsilon$. We find that the magnetic fields reconstructed using the Curlometer have a total divergence of $13.40 \epsilon$, the RBF reconstructed fields have $5.78 \epsilon$, and the Timesync reconstructed fields have $4.98 \epsilon$. Because $\epsilon$ is assumed to be very small in a physical system, it appears that all three methods preserve the divergence-free property of the field. However, this calculation indicates that the Curlometer method results in magnetic fields which are more unphysical than those reconstructed via the RBF and Timesync methods.

\subsection{Point-Wise Errors}
\begin{figure}
\centering
\textbf{Point-Wise Errors}\par\medskip
\includegraphics[trim={0.45in 0.45in 0 0.6in}, clip, width=1\textwidth]{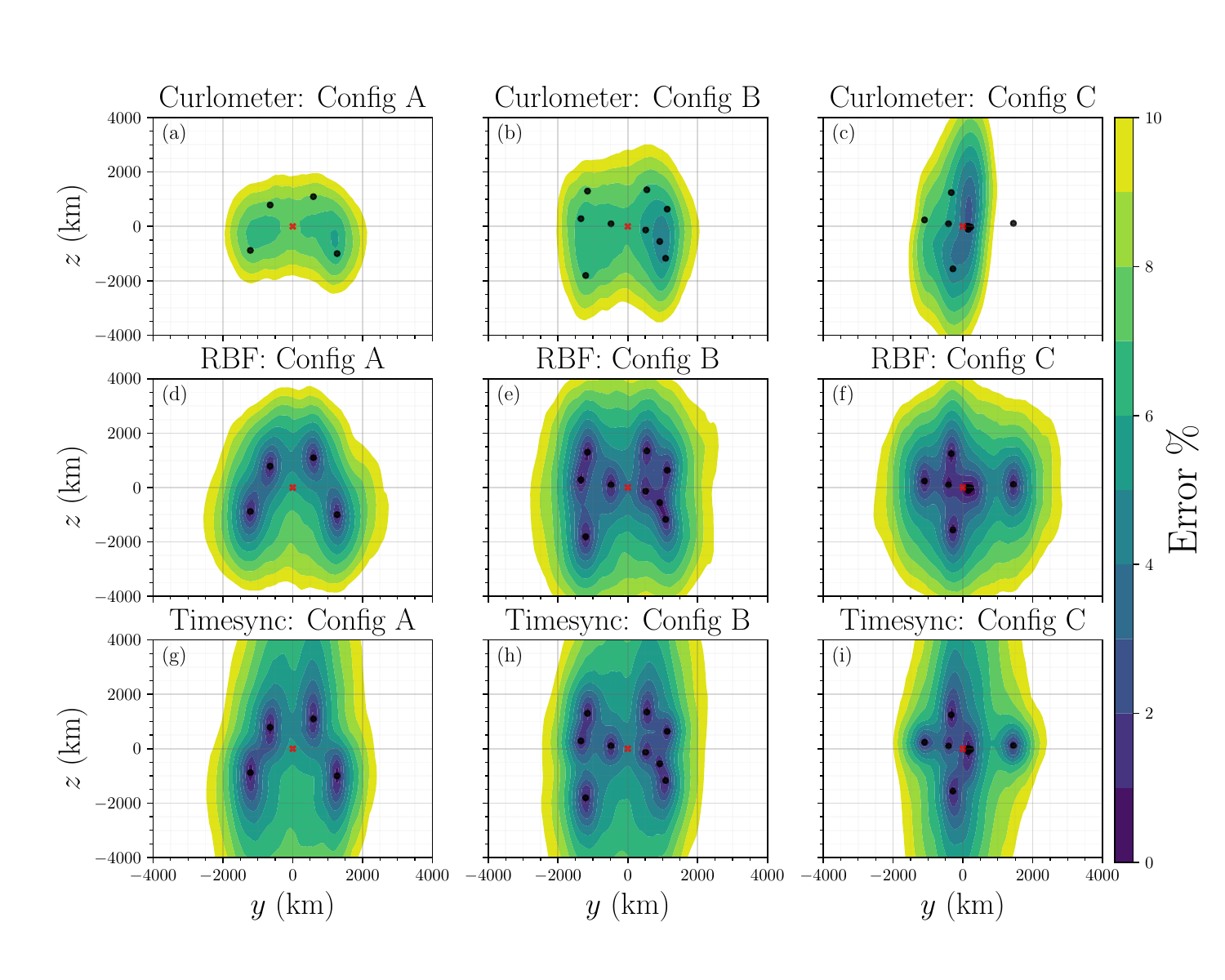}
%trim={<left> <lower> <right> <upper>}
\caption{The spatial distribution of the average (over both the out of plane direction and 20 spatially disjoint samplings) value of the point-wise error (Eqn \ref{eqn:Error_B}) around spacecraft configurations A (left column), B (middle column), and C (right column). We show results using the Curlometer method (first row), the RBF method (second row), and Timesync method (third row). The spacecrafts' projected positions are shown as black dots, and the spacecraft configuration barycenters as red xs.}
\label{fig:Brecon_error_yz}
\end{figure}

By averaging over the 20 disjoint realizations through the plasma simulation, we gather the average error with respect to an observatory position over the $56$ second ($4$ Hz) sampled 3D volume. We then take the arithmetic mean (the mean and median were verified to be similar) along the direction of spacecraft trajectory $\hat{x}$ to compute the 2D figures of average point-wise error in the $yz$-plane in Figure \ref{fig:Brecon_error_yz}.

It is clear from this figure that the Curlometer method performs the worst for all configurations. The interpolation using this method both has the smallest region of $<10\%$ error and does not exactly match the spacecraft observations at the spacecraft locations. To further investigate the differences in topological reconstructions found using these methods, we compute what fraction of the area within a circle of radius $2L$ is reconstructed with varying levels of precision. The results of this computation are shown in Table \ref{tab:error_vols}.

\begin{table}
\centering
    \begin{tabular}{ll|lll}
    Method    & Error Threshold  &  Config A     &  Config B    &  Config C    \\ \hline  
    Curlometer  & 20\%          & 0.954        & 0.981       & 0.674 \\
                & 15\%          & 0.688        & 0.805       & 0.502 \\
                & 10\%          & 0.285        & 0.428       & 0.311 \\
                & 5\%           & 0.002        & 0.021       & 0.064 \\
                & 1\%           & 0.000        & 0.000       & 0.000 \\ \hline
    RBF         & 20\%          & 1.000        & 1.000       & 1.000 \\
                & 15\%          & 0.982        & 0.997       & 0.994 \\
                & 10\%          & 0.643        & 0.742       & 0.711 \\
                & 5\%           & 0.139        & 0.247       & 0.196 \\
                & 1\%           & 0.004        & 0.012       & 0.013 \\ \hline
    Timesync    & 20\%          & 1.000        & 1.000       & 1.000 \\
                & 15\%          & 0.986        & 0.990       & 0.966 \\
                & 10\%          & 0.668        & 0.683       & 0.571 \\
                & 5\%           & 0.192        & 0.249       & 0.170 \\
                & 1\%           & 0.006        & 0.009       & 0.006 \\
    \end{tabular}

    \caption{Fraction of region within a distance of $2L$ (where $L=2000$ km) from the three spacecraft barycenters that are reconstructed with an error of less than 20, 15, 10, 5, and 1\% for each of the spacecraft configurations and three reconstruction methods.}
    \label{tab:error_vols}
    \vspace{-1cm}
\end{table}

This table again supports the conclusion from Figure \ref{fig:Brecon_error_yz} that a polynomial approximation approach, such as is used in our Curlometer method, yields the most mismatched reconstructed magnetic fields. The RBF and Timesync methods perform very similarly, as the Timesync seems to be slightly better for configuration A while the RBF seems to perform slightly better for configurations B and C.

However, we see in Figure \ref{fig:Brecon_error_yz} that if the region of comparison was extended further along the $\hat{z}$ direction, the Timesync method would yield lower errors at further distances. The sharp decline in performance of the RBF method in the domain outside of the spacecraft configuration is a known phenomena. It has previously been attributed to both the effects of Runge's phenomenon \cite{Runge:1901} and other mechanisms known to decrease the accuracy of RBF reconstructions near boundaries \cite{Fornberg:2002}.

We suspect that the poor performance of the Curlometer method is due to two factors.  First, we implemented the Curlometer to reconstruct a plane of points on the $yz$-plane which passes through the configuration's barycenter. Because the spacecraft did not take measurements on the $yz$-plane at the time of reconstruction, we also do not expect that the measured values of $\B$ and the reconstructed values of $\B$ will match identically at the projected $yz$ spacecraft positions. Additionally, when the Curlometer is applied to configurations of more than four spacecraft \cite{Broeren:2021}, the interpolated field, because it is a composite of many estimates, no longer exactly matches the spacecraft observations at the locations of the observatories. The offset and statistical composite nature of the $N > 4$ Curlometer method explains the lack of volume where the error less than 1\%.

Second, the Curlometer computes the spatial derivatives of the magnetic field as constant with respect to spatial position. This constraint leads to magnetic field reconstructions which vary linearly in space. While this assumption is a necessity if one desires to compute current density using the spatial derivatives, it is an unrealistic constraint to be imposed on the magnetic field of a turbulent plasma. Conversely, the RBF and Timesync methods directly interpolate the magnetic field. This direct interpolation does not attempt to compute or constrain the spatial derivatives of the field, and therefore has more degrees of freedom to create a reconstructed field which more closely resembles the fields sampled by the spacecraft.

\subsection{Wasserstein Distances}
For each spacecraft configuration and reconstruction method combination, we compute the relative Wasserstein distance $\overline{W}_d$ between each reconstructed $yz$-plane and the baseline truth from the simulation. This computation is done for the region of space within a circle of radius $L=2000$ km. For all of these combinations, we compute $\overline{W}_d$ separately for $B_x$, $B_y$, and $B_z$ components.

In the right-hand panels of Figure \ref{fig:Bx_compare}, the relative Wasserstein distance of each example $yz$-plane from the simulation is computed. In the left-hand panels, we report the 10-90 percentile ranges of relative Wasserstein distances that were computed using all 225 $yz$-planes, collocated along the $x$-axis. We see for this field, the RBF method has the lowest values of $\overline{W}_d$ and also produces the reconstructed $B_x$ component of the field that appears to best visually match the simulation. This test is verification that the relative Wasserstein distance is a reasonable quantification of topologically similarity.

\begin{figure}
\centering
\textbf{Relative Wasserstein Distances}\par\medskip
\includegraphics[trim={0.5in 0.5in 0.7in 0.6in}, clip, width=1\textwidth]{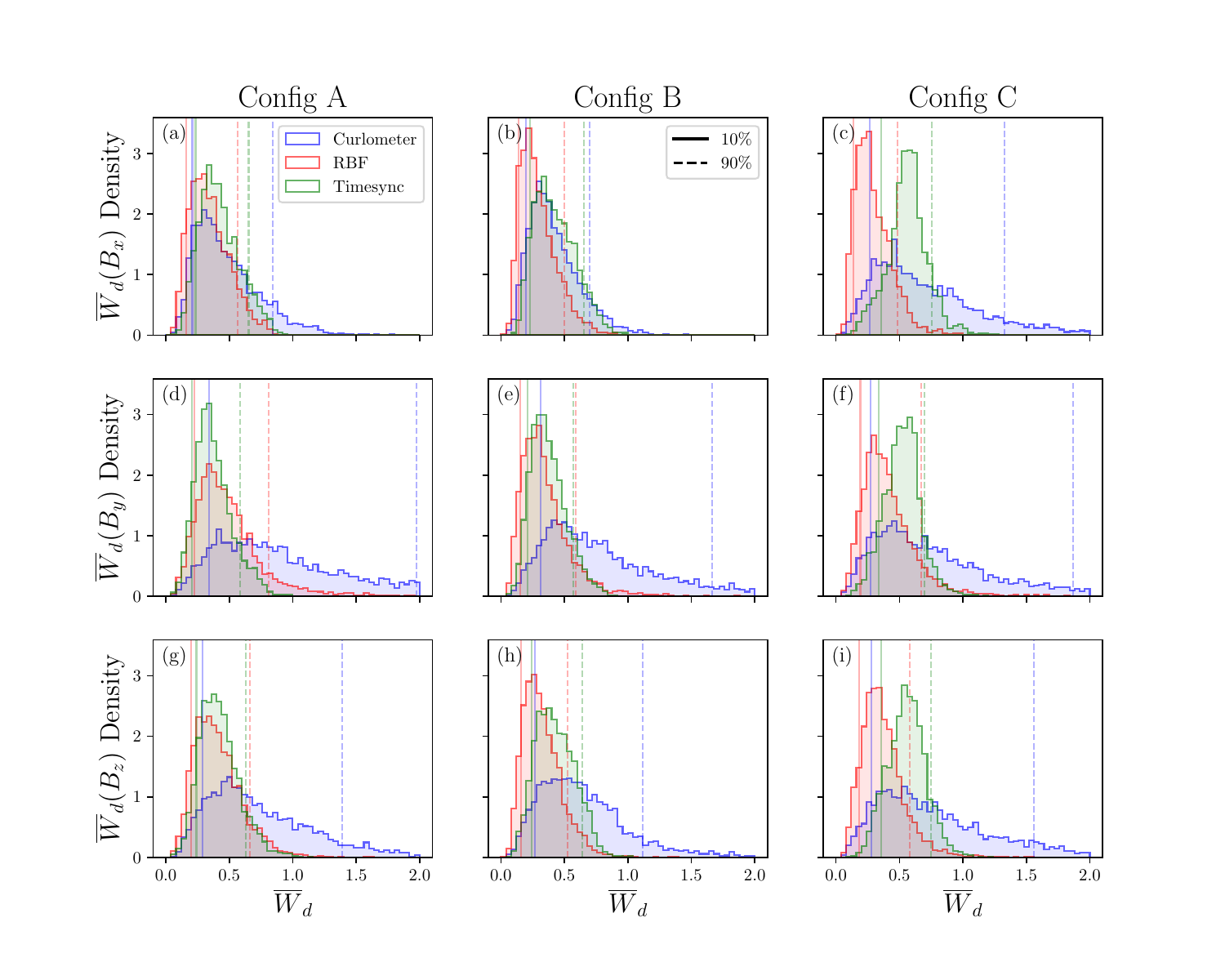}
%trim={<left> <lower> <right> <upper>}
\caption{Distributions of relative Wasserstein distances for each slice in the reconstructed $yz$-plane vs the simulated plane using the three different reconstruction methods (colors). The distributions of the $B_x$ (top row), $B_y$ (middle row), and $B_z$ (bottom row) magnetic fields were compared independently for each spacecraft configuration (columns).  Each distribution is normalized to have unity area. The vertical solid and dashed lines represent the 10\% and 90\% levels of each distribution respectively. }
\label{fig:W_distance_dists_normalized}
\end{figure}

In Figure \ref{fig:W_distance_dists_normalized}, we show the distribution of relative Wasserstein distances using each $yz$-plane of the reconstructions. From this figure, we see that the Curlometer has the largest spread and magnitude of $\overline{W}_d$ values. This spread implies that the topology of the magnetic field the Curlometer produces contains large, non-constant inconsistencies with the simulated fields. By counting the number of instances where $\overline{W}_d \geq 1$ for each reconstruction method, we find that a spatially constant reconstruction will outperform the Curlometer in $21.9\%$ of cases, outperform the RBF in $1.3\%$ of cases, and outperform the Timesync in $0.2\%$ of cases.

From all of the distributional data in Figure \ref{fig:W_distance_dists_normalized}, we compute the median value of $\overline{W}_d$ for each component, configuration, and method combination and display the results in Table \ref{tab:Wass_vals}. By examining the rows of this table, we find that (for the median case) the RBF method produces the field with the most topological similarity to the true field in 8 of the 9 spacecraft configuration and magnetic field component cases.

\begin{table}
\centering
    \begin{tabular}{ll|lll}
    Method      & Component     & Config A      & Config B     & Config C    \\ \hline  
    Curlometer  & $B_x$         & 0.421        & 0.382       & 0.614 \\
                & $B_y$         & 0.847        & 0.709       & 0.702 \\
                & $B_z$         & 0.664        & 0.590       & 0.694 \\\hline
    RBF         & $B_x$         & 0.322        & 0.263       & 0.266 \\
                & $B_y$         & 0.437        & 0.308       & 0.372 \\
                & $B_z$         & 0.381        & 0.300       & 0.342 \\ \hline
    Timesync    & $B_x$         & 0.403        & 0.405       & 0.565 \\
                & $B_y$         & 0.354        & 0.355       & 0.527 \\
                & $B_z$         & 0.403        & 0.415       & 0.557 \\
    \end{tabular}
    \caption{Median relative Wasserstein distances (Eqn \ref{eqn:Wd_tilde}) for each spacecraft configuration (columns A, B, C) and magnetic field reconstruction method as a function of each magnetic field component $B_x$, $B_y$, $B_z$ (rows).}
    \label{tab:Wass_vals}
    \vspace{-1cm}
\end{table}

\subsection{Statistical Fluctuation Distributions}
\begin{figure}
\centering
\textbf{Turbulent Fluctuations: Positional Lag in $x$}\par\medskip
\includegraphics[width=1\textwidth]{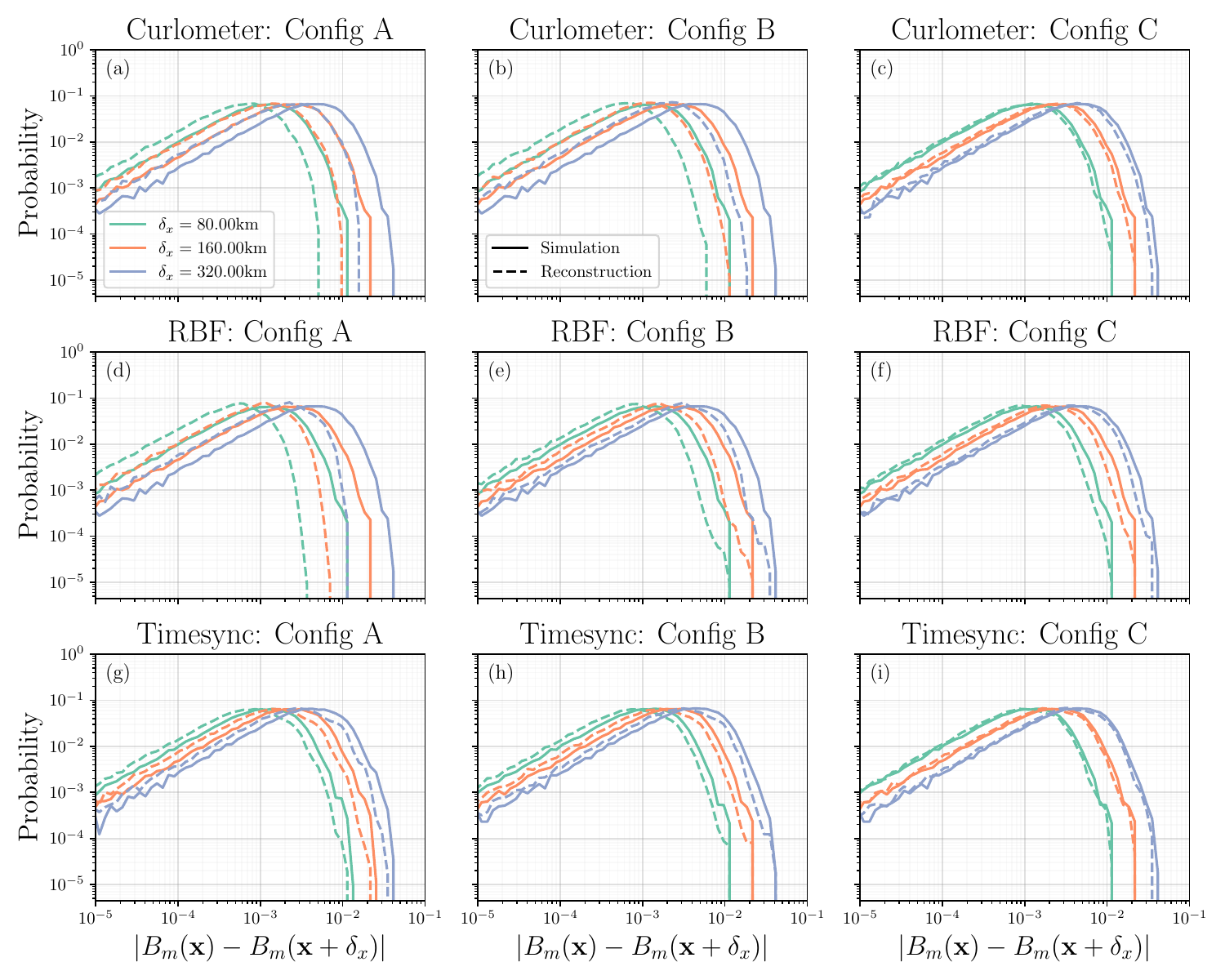}
\caption{Distribution of magnetic field fluctuations (Eqn \ref{eqn:B_dist}) for spacecraft configuration A (left column), B (middle column), and C (right column). We show results using the Curlometer method (first row), the RBF method (second row), and Timesync method (third row). The solid lines correspond to the baseline truth of the simulation, the dashed lines represent the fluctuations found in the corresponding reconstructed magnetic field. The lag distances in the $x$-direction, $\delta_x$, for the three values shown is displayed in the legend of panel (a).}
\label{fig:Bdist_x}
\end{figure}

\begin{figure}
\centering
\textbf{Turbulent Fluctuations: Positional Lag in $y$}\par\medskip
\includegraphics[width=1\textwidth]{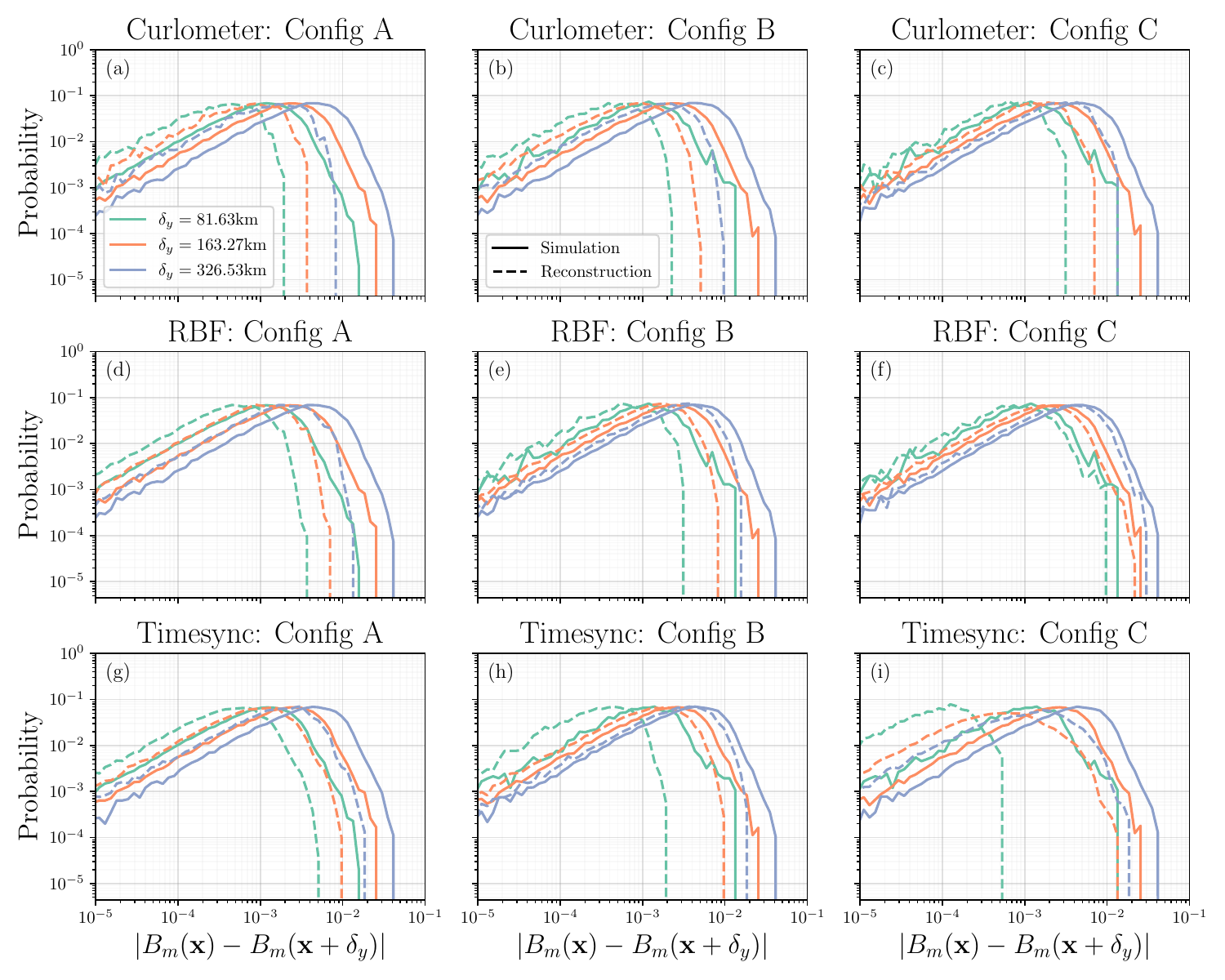}
\caption{Distributions of magnetic field fluctuations (Eqn \ref{eqn:B_dist}) for spacecraft configurations and reconstruction methods at three different lag distances in the $y$-direction, $\delta_y$, plotted in the same format as Figure \ref{fig:Bdist_x}.}
\label{fig:Bdist_y}
\end{figure}

We also investigate whether the fields reconstructed via the reconstruction methods preserve the statistical properties of the turbulence that they are trying to emulate. As distributions of scale dependent fluctuations are a hallmark property of turbulence, we compare the magnetic field fluctuations found in the reconstructed field to the true values, directly extracted from the turbulence simulation. We compute the fluctuations of the three magnetic field components via Eqn \ref{eqn:B_dist} and plot a histogram of the values in Figures \ref{fig:Bdist_x} and \ref{fig:Bdist_y}. These figures display the distributions at three lag distances, in the $\hat{x}$ and $\hat{y}$ directions respectively. The $\delta$ scales plotted in these figures represent 1, 2, and 4 times the $x$ and $y$ minimum resolution of the reconstructed field. In these panels, a perfect reconstruction would have overlapping dashed and solid lines of the same color.

Figure \ref{fig:Bdist_x} shows that the Timesync method captures the distributions of magnetic field fluctuations in the $\hat{x}$ direction for all configurations, while the Curlometer and RBF perform better or worse depending on the configuration of spacecraft. This result suggests that along the direction of travel, the Timesync method is more resilient to variations in spacecraft configuration than the other methods. This figure also shows that all three of the methods are able to accurately reproduce the fluctuations in the $\hat{x}$ direction for spacecraft configuration C. Because the Curlometer and RBF produced offset distributions for configuration B but not C, we conclude that number of spacecraft is not the only important factor that contributes to these methods' accuracy.

Figure \ref{fig:Bdist_y} shows that all of the methods perform worse in the direction perpendicular to the spacecraft direction of travel. All of the methods still produce scale-dependent structure; however, the distributions do not match those gathered from the simulation. Perhaps more disturbingly, the offset from true to reconstructed distribution is not consistent across scales for all of the methods. The only method and configuration combination which is a good reproduction is the RBF method using configuration C. Further investigation will be required to determine what geometric properties of spacecraft configuration C are leading to good statistical RBF performance in the $\hat{y}$ direction.

\subsection{Structure Functions}
\begin{figure}
\centering
\textbf{Structure Functions: Positional Lag in $x$}\par\medskip
\includegraphics[width=1\textwidth]{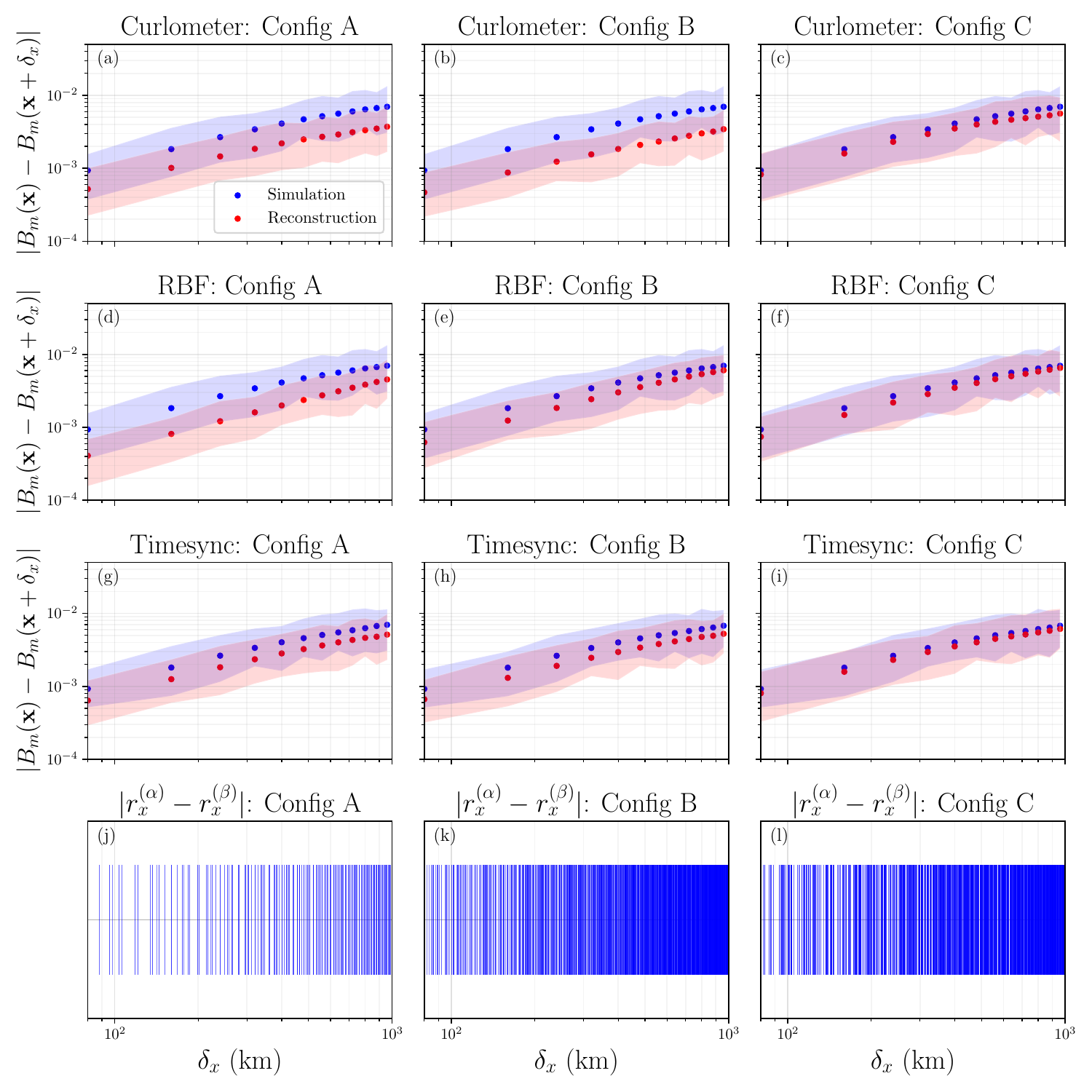}
\caption{Structure functions as a function of scale (Eqn \ref{eqn:S_1}) for spacecraft configuration A (left column), B (middle column), and C (right column). We show results using the Curlometer method (first row), the RBF method (second row), and Timesync method (third row). The blue dots are the median values from the simulation and the red dots are the median values from the reconstructed fields. The shaded region of each color represents the 25\%-75\% range. Scales range from grid scale (80km) to near half the spacecraft configuration scale ($L/2=1000$km). Panels (j)-(l) are spike raster plots showing the $x$ component of the inter-spacecraft distances for each configuration.}
\label{fig:S1_x}
\end{figure}

\begin{figure}
\centering
\textbf{Structure Functions: Positional Lag in $y$}\par\medskip
\includegraphics[width=1\textwidth]{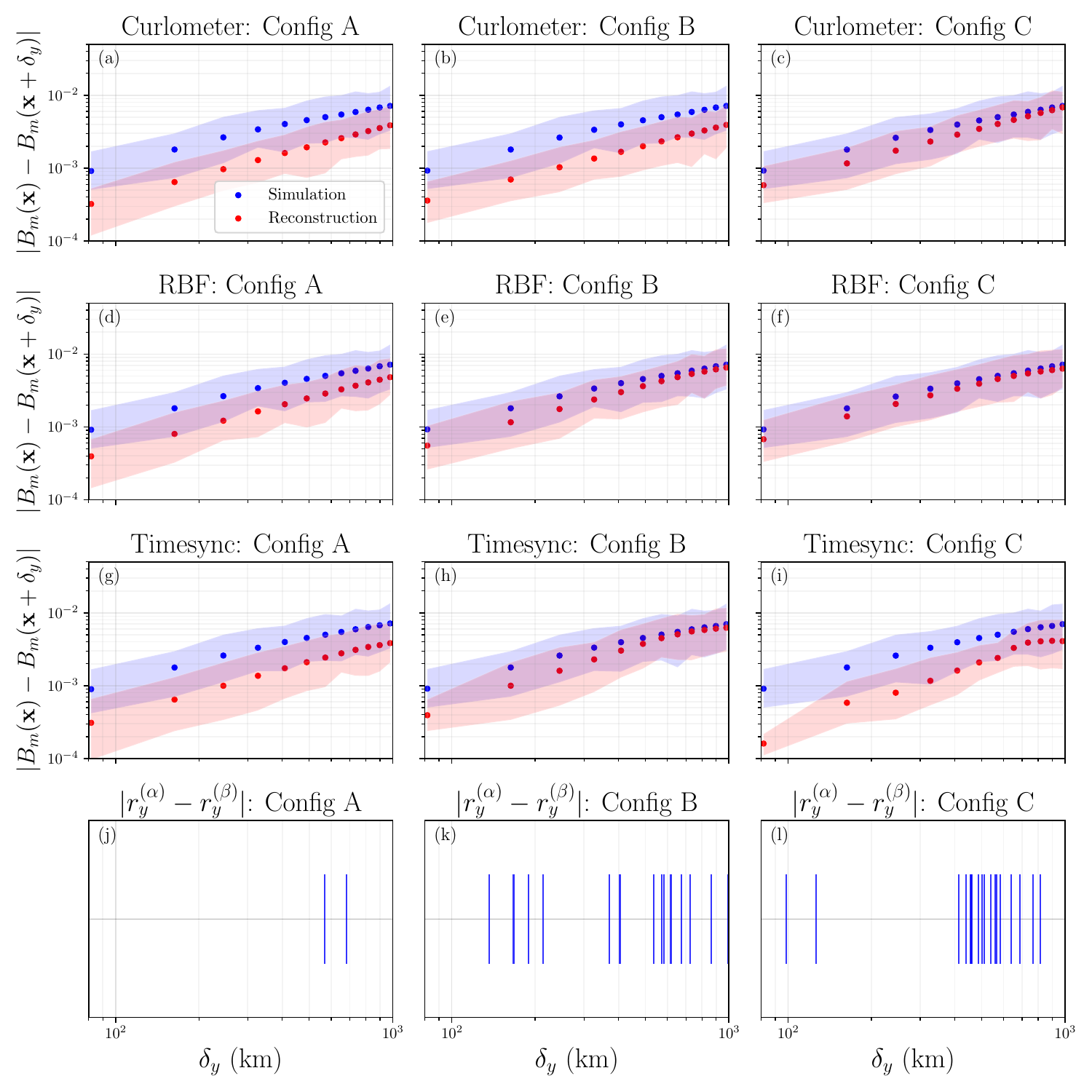}
\caption{Structure functions and inter-spacecraft distances, as a function of scale $\delta_y$, formatted in the same manner as Figure \ref{fig:S1_x}.}
\label{fig:S1_y}
\end{figure}

To analyze the distribution of fluctuations across a broader range of scales, we plot the first order structure function (Eqn \ref{eqn:S_1}) of the reconstructed magnetic field with the ground-truth value derived from the plasma simulation in Figures \ref{fig:S1_x} and \ref{fig:S1_y}. We find that the Timesync method matches the ground truth most closely for all three configurations at all scales $\delta_x \in [80,1000]$ km. We also observe that all of the methods match the ground truth closely for spacecraft configuration C when using $x$ positional lags.

We analyze the average fluctuations in the $\hat{y}$ direction across the scales $\delta_y \in [80,1000]$ km in Figure \ref{fig:S1_y}. We see that the RBF method applied to configuration C is accurate across a wide range of scales, while all other combinations are not. Interestingly, the Timesync method does not follow a near power-law scaling for configurations B and C, as the fluctuations appear to be stair-stepped. By plotting the differences in the $y$ components of the spacecraft positions in the bottom row of Figure \ref{fig:S1_y}, we see that this is likely due to the lack of spacecraft separations $\delta_y$ in the range $[126,413]$ km for spacecraft configuration C. Conversely, this stair-stepping phenomenon was not observed for the structure functions with positional differences $\delta_x$. From the bottom row of Figure \ref{fig:S1_x}, we conclude that this phenomenon is likely because there are no gaps in the $x$ component of the inter-spacecraft positions.

\begin{figure}
\centering
\textbf{Structure Function Scaling Exponent}\par\medskip
\includegraphics[width=1\textwidth]{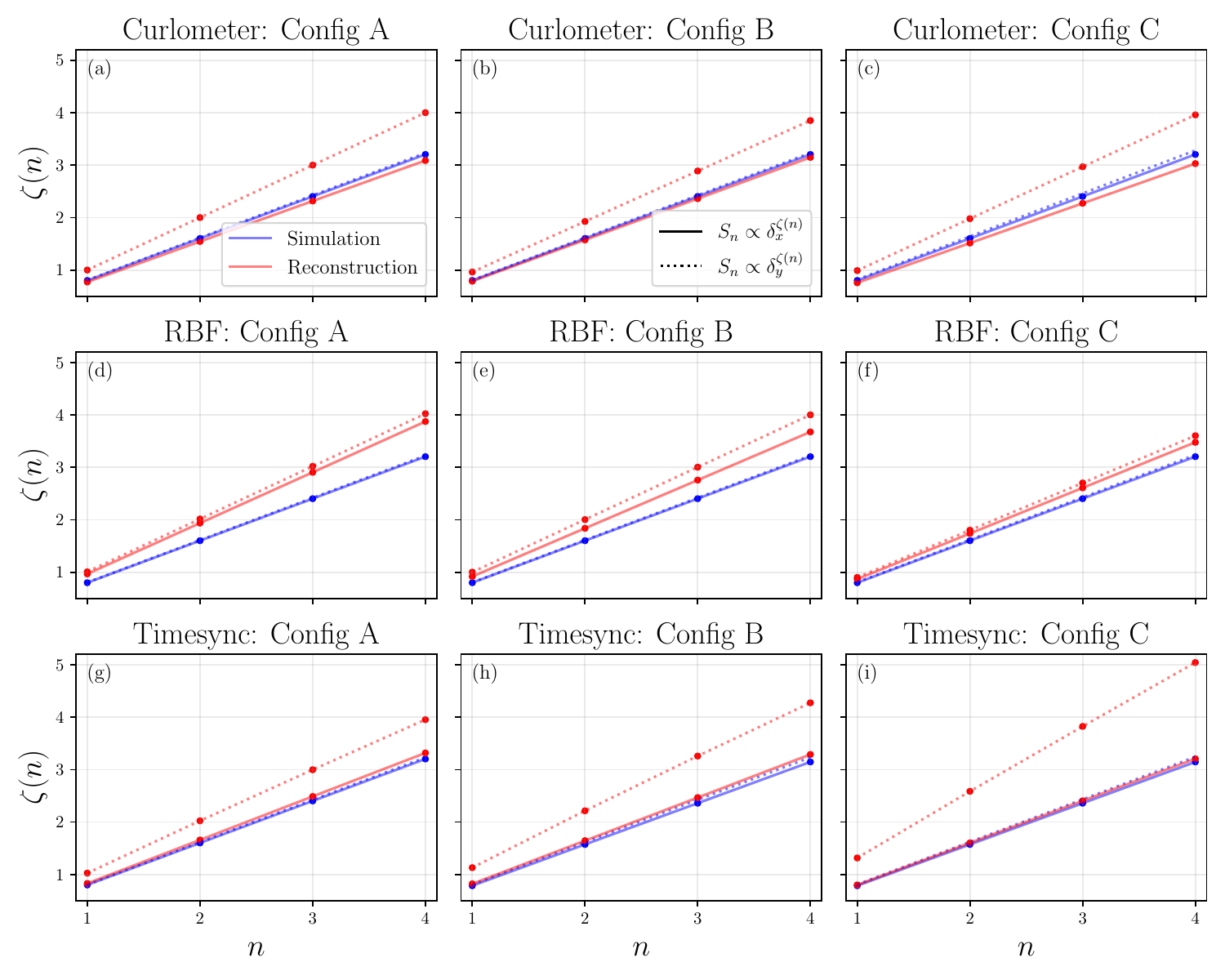}
\caption{For each spacecraft configuration (columns) and reconstruction method (rows), we compute the scaling coefficient of the structure functions $S_n$ ($n \in \{1,2,3,4\}$) assuming $S_n \propto \delta^{\zeta(n)}$. The solid lines represent the functions $S_n$ derived using positional differences $\delta_x$ and the dashed lines represent the structure functions derived using $\delta_y$ positional differences. The color represents if the data came from the simulated (blue) or the reconstructed (red) field.}
\label{fig:zeta}
\end{figure}

The scaling exponent $\zeta(n)$ of the structure functions are often analyzed in plasma turbulence as a function of order $n$. We fit functions of the form $S_n(\delta) \propto \delta^{\zeta(n)}$ to the first four structure functions, treating the $\delta_x$ and $\delta_y$ lags separately, and plot the results in Figure \ref{fig:zeta}. We analyze only four orders of structure functions $n$ because the maximum order that can be resolved is related to the number of points in the dataset through the formula $n_{max}=\log_{10}(M)-1$ \cite{dudok:2004, dudok:2013}; for our dataset,  $M=445500$, and thus $\log_{10}(M)-1\approx 4.65$.

We find that across all methods and configurations, the scaling exponent is most incorrect for lags in the $\delta_y$ direction. The Timesync method accurately captures $\zeta(n)$ for all configurations in the $\delta_x$ direction, while the other two methods do not. The configuration and reconstruction method combination that appears most accurate in capturing $\zeta(n)$ for $\delta_x$ and $\delta_y$ is the RBF method applied to configuration C.

Across Figures \ref{fig:S1_x} to \ref{fig:zeta}, we have observed the Timesync method both performing the best in the $\hat{x}$ direction, and the worst in the $\hat{y}$ direction. We believe that the large differences seen in the Timesync performance are due to how it asymmetrically treats the $x$ (parallel/nearest-neighbor) and $y$ (perpendicular/2D IDW) directions. The Curlometer and RBF methods do not vary interpolation schemes along these two directions, and therefore have more consistent statistical performances.

\section{Conclusions}
\label{sec:conclude}
We have performed a comparison of three multi-spacecraft magnetic field reconstruction techniques: Curlometer, RBF, and Timesync. The Curlometer is an established technique for reconstructing magnetic fields using multi-spacecraft data, while the RBF and Timesync methods are more novel approaches that we have defined and analyzed. We note that the computational complexity of the Curlometer method scales super-exponentially with number of spacecraft that compose an observatory, while the RBF and Timesync scale linearly. 

For large-scale topological reconstructions of time-varying turbulent magnetic fields, we have shown that the Curlometer approach performs far worse than the other two methods. As it does not preserve a perfect reproduction of the observed data near positions of the spacecraft, it is unable to reproduce any volume of space with magnetic fields with less than 1\% error on average. Also, it only reconstructs about half as much volume within a distance of $2L$ from the spacecraft configurations barycenter with $\leq 10\%$ error, when compared to the other two methods. By combining the point-wise error results with the relative Wasserstein distance metric, we can conclude that the Curlometer is the least likely method to preserve the topological structures found in the measured magnetic field.

We examined the small-scale statistical fluctuations of the reconstructed magnetic fields and compared them to the simulation of turbulence that they were attempting to reproduce. We found that all three of the methods were capable of reproducing scale-dependent distributions of fluctuations, in the form of structure functions, that are a signature of turbulence. However, the absolute value of these functions was shifted by various factors for the different spacecraft configuration and reconstruction method combinations. The Timesync method was the most accurate method to reproduce fluctuations along the spacecraft direction of travel, while the RBF method was the most accurate option to reproduce fluctuations perpendicular to the spacecrafts' travel direction. We also found that fluctuations along both directions were easiest to reproduce when using our HelioSwarm-like configuration of nine spacecraft.

In totality, our results suggest that a direct interpolation approach (which does not estimate spatial derivatives), such as our RBF or Timesync methods, has the most potential to reconstruct accurate magnetic fields from sparse in situ multi-spacecraft data. A further study of the RBF and Timesync methods as a function of the number of spacecraft, shape of configuration, and direction of spacecraft travel is needed to quantify the uncertainty in these methods. Such studies would ideally use time-varying turbulent magnetic field data and include many of the uncertainty quantification techniques outlined in this work.

\section{Data Availability}
Codes demonstrating each of the three magnetic field reconstruction methods can be found at the corresponding author's personal GitHub repository \url{https://github.com/broeren/B_Field_Reconstructions}. This work was developed in Python 3.8.11 using the numpy \cite{numpy:2020}, scipy \cite{SciPy:2020}, and matplotlib \cite{matplotlib:2007} packages extensively.

We make use of data from a time-varying simulation of numerical turbulence, designed by Dr. Jason TenBarge. \gkeyll\ is open source and can be installed by following the instructions on the \gkeyll\ website (http://gkeyll.readthedocs.io). The input file for the \gkeyll\ simulation presented here is available in the following GitHub repository, https://github.com/ammarhakim/gkyl-paper-inp. 

%Authors should include an Availability Statement for the software that has a significant impact on the research. Details and templates are in the Availability Statement section of the Data and Software for Authors Guidance: \url{https://www.agu.org/Publish-with-AGU/Publish/Author-Resources/Data-and-Software-for-Authors#availability}

%%%%%%%%%%%%%%%%%%%%%%%%%%%%%%%%%%%%%%%%%%%%%%%

\acknowledgments
Funding for phase B of the HelioSwarm mission was provided under NASA contract no. 80ARC021C0001. K.G.K was supported by NASA Early Career Grant 80NSSC19K0912. This research was supported by the International Space Science Institute in Bern, through ISSI International Team project \#556 (\url{https://teams.issibern.ch/energtransferspaceplasmas/}) Cross-Scale Energy Transfer in Space Plasmas. J.M.T. was supported by the NSF under grant number AGS-1842638. We thank Dr. Colby Haggerty for his insights during discussions with us.

%% ------------------------------------------------------------------------ %%
%% References and Citations            
%%%%%%%%%%%%%%%%%%%%%%%%%%%%%%%%%%%%%%%%%%%%%%%

\bibliography{main}

\appendix
\section{Scale Independence of Curlometer}
\label{appendix:sec:curl_scale}
The scale-dependent magnetic field fluctuations that we wish to analyze are defined as
\begin{equation}
	\left| \B_m(\mathbf{r}) - \B_m(\mathbf{r} + \lambda) \right|.
\end{equation}
The magnetic field reconstruction of the Curlometer is based on a first-order Taylor series
\begin{equation}
	\B_m(\mathbf{r}^{(i)}) = \B_m(\mathbf{r}) + \sum_{k \in \{x,y,z\}} \partial_k \B_{m} \left(r^{(i)}_k - r_k \right).
\end{equation}
If we rearrange this equation, we see that 
\begin{equation}
	\B_m(\mathbf{r}) = \B_m(\mathbf{r}^{(i)}) - \sum_{k \in \{x,y,z\}} \partial_k \B_{m} \left( r^{(i)}_k - r_k \right). \label{eqn:appendix_sub}
\end{equation}
We can substitute the point $\mathbf{r} + \lambda$ for $\mathbf{r}$ so that we have
\begin{align}
	\B_m(\mathbf{r} + \lambda) &= \B_m(\mathbf{r}^{(i)}) - \sum_{k \in \{x,y,z\}} \partial_k \B_{m} \left( r^{(i)}_k - (r_k + \lambda_k) \right) \\
	\B_m(\mathbf{r} + \lambda) &= \left[ \B_m(\mathbf{r}^{(i)}) - \sum_{k \in \{x,y,z\}} \partial_k \B_{m} \left( r^{(i)}_k - r_k\right) \right]  +  \sum_{k \in \{x,y,z\}} \partial_k \B_m \lambda_k
 \end{align}
 Substituting Eqn \ref{eqn:appendix_sub} into the brackets of the above equation, we find that
 \begin{equation}
     \B_m(\mathbf{r} + \lambda) = \B_m(\mathbf{r}) +\sum_{k \in \{x,y,z\}} \partial_k \B_m \lambda_k.
 \end{equation}
Therefore, we have
\begin{equation}
	\left| \B_m(\mathbf{r} + \lambda) - \B_m(\mathbf{r}) \right| = \left| \sum_{k \in \{x,y,z\}} \partial_k \B_m \lambda_k \right|.
\end{equation}
However, the matrix $\partial \B$ is a computed constant in this formulation, and does not vary as a function of spatial location $\mathbf{r}$. This fact means that a reconstruction using the Curlometer does not have a distribution of fluctuations, but is instead given by a linear relationship with $\lambda$
\begin{equation}
	\left| \B_m(\mathbf{r} + \lambda) - \B_m(\mathbf{r}) \right| \sim  \lambda.
\end{equation}
Using more than 4 spacecraft, we can combine estimates which individually have this property. An arithmetic mean will preserve the linear dependence property as any linear combination of terms linear in $\lambda$ will remain linearly dependent on $\lambda$. We therefore choose to combine groups of Curlometer estimations using the component-wise median, which cannot be expressed as linear combinations.

\end{document}